\documentclass[5p,sort&compress]{elsarticle}
\usepackage{amssymb}
\usepackage{amsmath}
\usepackage{color}
\usepackage{graphicx}
\usepackage{caption}
\usepackage{subcaption}
\usepackage[english]{babel}

\newcommand{\beq}{\begin{eqnarray}}
\newcommand{\eeq}{\end{eqnarray}}

\begin{document}

\begin{frontmatter}

\title{Proton-$\eta^\prime$ interactions at threshold}

\author[label1,label2]{A.~V.~Anisovich}
\author[label3]{V.~Burkert}
\author[label4]{M.~Dugger}
\author[label1,label3]{E.~Klempt}
\author[label1,label2]{V.~A.~Nikonov}
\author[label4]{B.~G.~Ritchie}
\author[label1,label2]{A.~Sarantsev}
\author[label1]{U.~Thoma}

\address[label1]{Helmholtz--Institut f\"ur Strahlen-- und Kernphysik, Universit\"at Bonn, Germany}
\address[label2]{NRC ``Kurchatov Institute'', PNPI, Gatchina 188300, Russia}
\address[label3]{Thomas Jefferson National Accelerator Facility, Newport News, Virginia 23606}
\address[label4]{Arizona State University, Tempe, Arizona 85287-1504}

\begin{abstract}
Recent data on photoproduction of $\eta'$ mesons off protons have been included in the 
data base for the Bonn-Gatchina partial wave analysis. The real and 
imaginary parts of the $S$-wave $\eta^\prime p\to \eta^\prime p$ scattering amplitude 
in the threshold region were fit to yield the $\eta ‘ p$ scattering length and the 
interaction range. This new analysis found $|a_{\eta' p}|= 
(0.403\pm 0.015\pm 0.060)$\,fm and a phase $\phi=(87\pm2)^\circ$, 
while the range parameter is not well-constrained. The striking behavior of the 
GRAAL data on the beam asymmetry in the threshold region suggests that a narrow
proton-$\eta'$ resonance might exist. However, the scattering length was found to be relatively 
insensitive to the possible existence of this narrow resonance.
\end{abstract}

\begin{keyword}
baryon spectroscopy  \sep meson photoproduction \sep polarization observables 
\end{keyword}

\end{frontmatter}
The interaction of the  $\eta'$ meson with the nucleon is a very active area of 
research at present, both in theoretical attempts to understand the interaction 
and in experiments aimed at providing polarization observables. The $\eta'$-meson  
is a member of the nonet of ground-state pseudoscalar mesons. Unlike other meson 
nonets, the $\eta'$-meson within the pseudoscalar nonet is nearly a pure SU(3) 
singlet state and may even contain contributions
from a pseudoscalar glueball \cite{Coffman:1988ve,Feldmann:1998vh,Escribano:2008rq,%
Cao:2012nj,Aaij:2014jna,Harland-Lang:2017mse}. The octet of pseudoscalar mesons plays
the role of Goldstone bosons; in the chiral limit of QCD when the quark masses go to
zero, their masses vanish as well. The topological structure of the QCD vacuum breaks 
the so-called $\rm U_A(1)$ symmetry, and the mass of the SU(3) singlet state, and hence
the $\eta'$ mass, does not vanish even for massless quarks. 

Nevertheless, a sizable fraction of the $\eta'$ mass is still due to chiral symmetry 
breaking. A key problem in non-perturbative QCD is wether the chiral symmetry can partially be
restored in a strongly interacting environment \cite{Bernard:1988db,Brown:1991kk,Hatsuda:1991ez}.
If so, then the partial restoration of chiral symmetry should lead to a reduction of the 
$\eta'$ mass, opening the possibility of the existence of $\eta' N$ bound states in 
nuclear matter. Even the existence of an $\eta’-$deuteron bound state in vacuum has 
been suggested~\cite{Sekihara:2017xkz}. A very recent determination of the $\eta'$-nucleus 
potential gave -- with carbon \cite{Nanova:2013fxl} or niobium \cite{Nanova:2016cyn} 
as nuclei -- a shallow potential of $-39\pm 7_{\rm stat}\pm15_{\rm syst}$\,MeV~\cite{Metag:2017yuh}
for $\eta'$-mesons with momentum of 1200 - 2900\,MeV/c. A new determination of the 
$\eta'$-nucleus potential with a mean $\eta'$ momentum of 600\,MeV/c is in progress~\cite{Nanova:2018tbd}.   
The scattering length and the range parameter for the $\eta'$-proton interaction 
also are directly related to the existence of any $\eta’ N$ bound states, as well.
In that regard, the $\eta'$-$p$ scattering length has been determined by
the COSY-11 Collaboration from the rise of the total cross section for the
reaction $p+p\to p+p+\eta'$ \cite{Czerwinski:2014yot}.

In this letter we report a determination of the S-wave $\eta' p$ 
length from an analysis of data on the reaction
\begin{eqnarray}
 \label{reac}
\gamma p\to \eta' p
\end{eqnarray}
This new analysis includes the recently-obtained data on the beam asymmetry 
$\Sigma$~\cite{Sandri:2014nqz,Collins:2017sgu} for the $\eta’$ photoproduction process, 
as well as the recent high-precision data on the differential cross for this
reaction~\cite{Kashevarov:2017kqb}. The data from GRAAL show a beam asymmetry that is  
larger near threshold ($W = 1896-1910$) than the value for that observable in the very next 
measured energy bin ($W=1910-1917$). The difference in the measured asymmetries is especially 
striking when one considers that the difference between the centers of those two energy 
bins is a mere 10\,MeV. In a partial-wave analysis, this remarkable behavior suggests
the existence of a narrow resonance at the $\eta' p$ threshold. 

\begin{figure*}[pt]
 \center
    \includegraphics[width=0.8\textwidth]{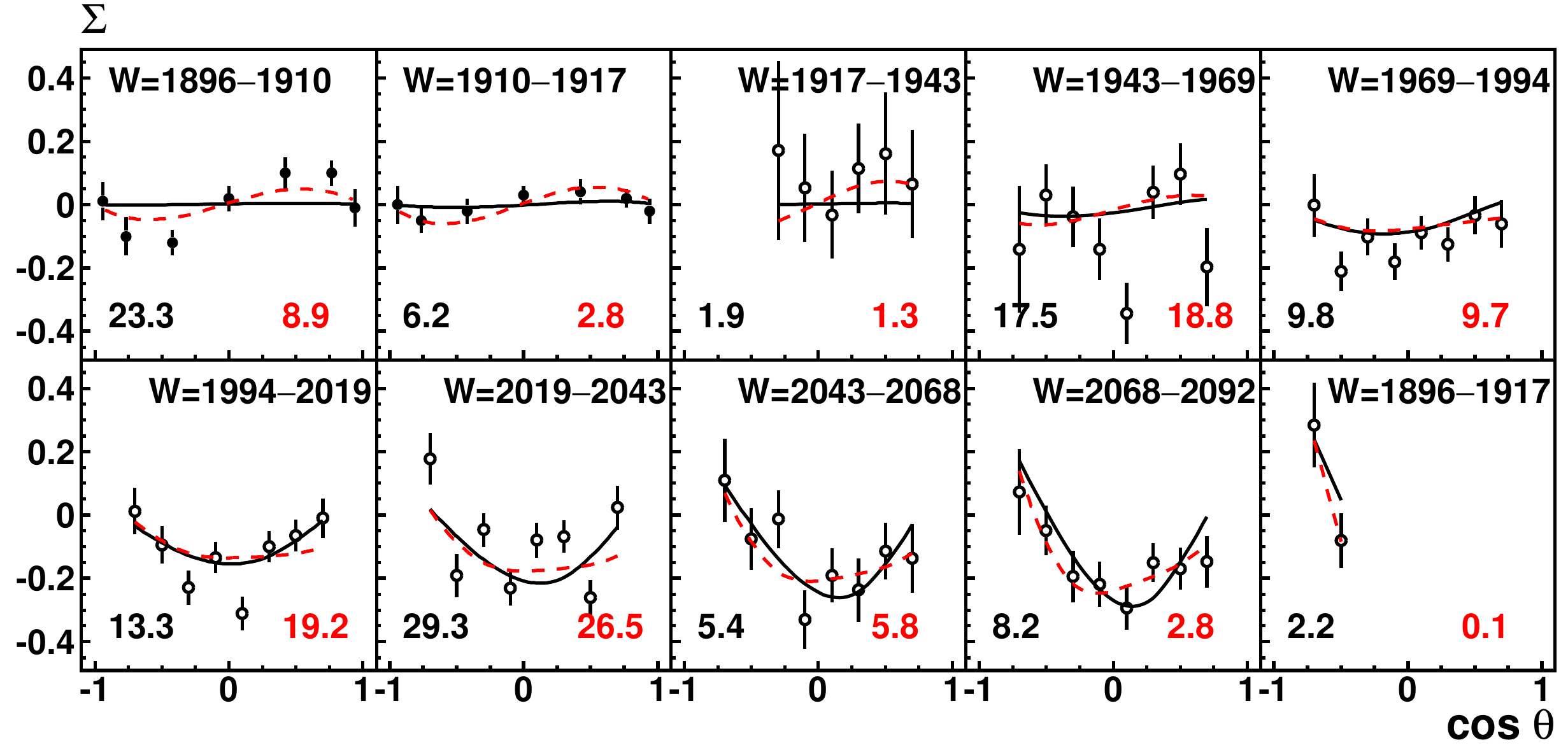}
  \caption{\label{asymmetry}(Color online)  The beam asymmetry $\Sigma$
for the reaction $\gamma p\to \eta' p$. Shown are data from GRAAL~\cite{Sandri:2014nqz}
(first two subfigures) and recent data from CLAS~\cite{Collins:2017sgu} (next eight subfigures).
The curves represent two fits: the solid (black) curve represents the main
fit without a new narrow resonance, the dashed (red) curve a fit which includes a narrow
$\eta'p$ threshold resonance with spin-parity $J^P=3/2^-$. The partial-wave analysis curves 
for the CLAS beam asymmetries are scaled by a factor 0.94 (see Ref.~\cite{Anisovich:2017tbd}).
}
\end{figure*}

Data on  the $\gamma p\to \eta' p$ reaction were analyzed recently with the aim of identifying 
the contributions to this reaction from different $N^*$ resonances, and to determine the
$N^*\to N\eta'$ branching ratios~\cite{Anisovich:2017tbd}. Four resonances,
$N(1895){1/2^-}$, $N(1900){3/2^+}$, $N(2100){1/2^+}$, and $N(2120){3/2^-}$, were found
to provide the most significant contributions. The fit used the differential cross
sections $d\sigma/d\Omega$ from the Crystal Barrel \cite{Crede:2009zzb} and
CLAS~\cite{Williams:2009yj} experiments, along with recent beam asymmetry data from GRAAL 
$\Sigma$ \cite{Sandri:2014nqz} and CLAS $\Sigma$~\cite{Collins:2017sgu} 
mentioned above. In the analysis presented here, we also used the new precise MAMI-A2
data on the $\gamma p\to \eta' p$ differential cross section~\cite{Kashevarov:2017kqb}.
In addition to the data on $\eta'$ photoproduction, a large body of pion and photo-induced reactions
is fit in a coupled channel analysis. The data base includes the real and
imaginary part of the $\pi N$ scattering amplitude for the partial waves up to $J=9/2^\pm$
from Ref.~\cite{Arndt:2006bf} and data on pion and photoproduction data with $\pi N$,
$\eta N$, $K\,\Lambda$, $K\,\Sigma$, $N\pi^0\pi^0$, and $N\pi^0\eta$ in the final state.
A list of the data with references can be found on our web page (pwa.hiskp.uni-bonn.de/).

Similar fits - even though mostly restricted to the $\gamma N\to \eta' N$ 
reaction - have recently been presented in 
Refs.~\cite{Zhong:2011ht,Huang:2012xj,Sakai:2016boo,Tryasuchev:2017amh}. 
In Ref.~\cite{Zhong:2011ht}, a quark model is used to fit differential cross 
sections for $\eta'$ photoproduction off protons~\cite{Williams:2009yj,Crede:2009zzb} 
and neutrons~\cite{Jaegle:2010jg}; Ref.~\cite{Huang:2012xj} includes data on
cross sections for $\pi^- p\to\eta'n$ and on $NN\to NN\eta'$ \cite{Klaja:2010vy}
(references to $\pi^-p$ and earlier $NN$ data can be found in Ref.~\cite{Huang:2012xj}). 
The authors of Ref.~\cite{Tryasuchev:2017amh} fit CLAS 2009 differential cross 
sections~\cite{Williams:2009yj} and the total cross section from  MAMI-A2 
collaboration~\cite{Kashevarov:2017kqb} within an isobar model. The 
analyses \cite{Anisovich:2017tbd,Zhong:2011ht,Huang:2012xj,Tryasuchev:2017amh}
agree that nucleon resonances should be included even though there is no consensus 
concerning the spin-parities of the preferred resonances. No explicit resonances 
were included in the analysis presented in Ref.~\cite{Sakai:2016boo}, but instead 
final-state interactions between $\eta'$ mesons and nucleons were studied within 
a three-flavor linear $\sigma$ model. 
The need of final-state interactions is demonstrated even though the quality of the fit is 
moderate. The recent CLAS data on the $\gamma p\to \eta'p$ beam asymmetry, which include a much wider energy range than that provided by GRAAL, \cite{Collins:2017sgu} have to date only been included in Ref.~\cite{Anisovich:2017tbd}.

The formalism used here to fit the data is described in Ref.~\cite{Anisovich:2017tbd}.
The fit solution presented in Ref.~\cite{Anisovich:2017tbd} predicted the new MAMI-A2 
data~\cite{Kashevarov:2017kqb} on the differential cross section for reaction~(\ref{reac}) reasonably
well; the inclusion of the data did not change the resonances contributing to $\eta'$
photoproduction, and the changes in their $N\eta'$ decay branching ratios were small.
Note that the errors given in~\cite{Anisovich:2017tbd} were dominantly due to a
variation of the model assumptions. The new MAMI-A2 data stabilize
the amplitudes close to the $\eta' p$ threshold.

\begin{figure}[pt]
 \center
    \includegraphics[width=0.48\textwidth,height=0.5\textheight]{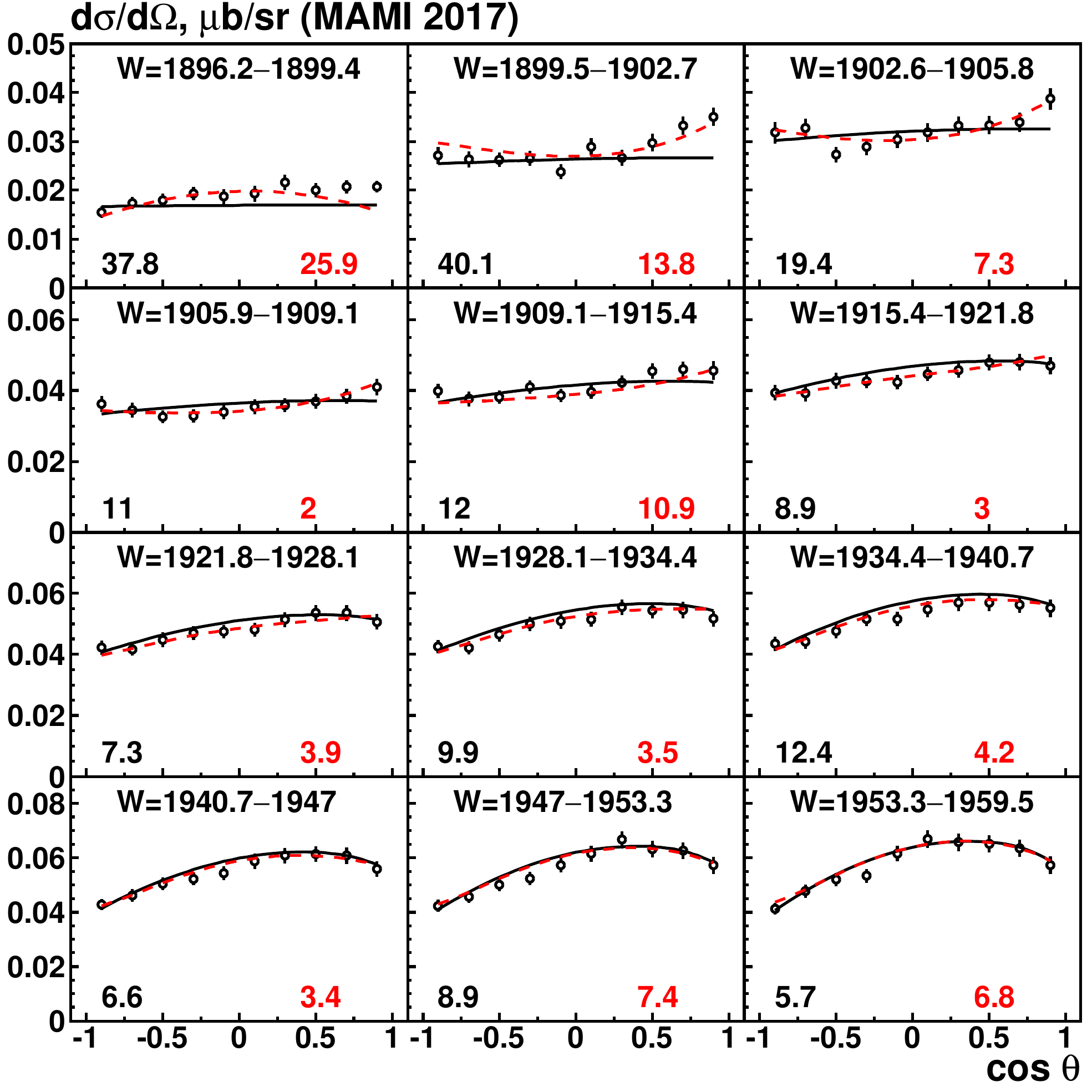}
  \caption{\label{etp_diff}(Color online) The MAMI-A2 differential cross section
for $\gamma p\to \eta' p$~\cite{Kashevarov:2017kqb} and two BnGa fits.
The dashed curve represents a fit which includes a narrow
$\eta'p$ threshold resonance with spin-parity $J^P=3/2^-$, the solid curve the main
fit without a narrow resonance.
}
\end{figure}

Figure~\ref{asymmetry} shows the GRAAL~\cite{Sandri:2014nqz} and CLAS~\cite{Collins:2017sgu} data
on the beam asymmetry $\Sigma$, while Fig.~\ref{etp_diff} gives differential cross section measured at 
MAMI-A2~\cite{Kashevarov:2017kqb}. The data are compared to two fits: 

1. Our {\it standard fit} is represented by the solid curves. This standard fit gives a 
reasonable description of the data except for the region just above the threshold. This  
fit predicts a vanishing beam asymmetry for the mass range where the GRAAL data exist, and
a relatively flat angular distribution for the differential cross section. However, in 
contrast to the predictions of this standard fit, the data GRAAL data show an appreciably 
beam asymmetry at threshold, and the new MAMI-A2 data indicate a significant forward 
rise of the differential cross section in the 1899.5 to 1902.7\,MeV mass range. 

2. We tried to improve the {\it standard fit} by including a narrow $N\eta'$ resonance. 
The narrow resonance was represented by a convolution of a squared Breit-Wigner amplitude and
a Gaussian function representing the resolution. The GRAAL resolution was fixed to 
$16$\,MeV (FWHM), for the MAMI-A2 data the resolution is expected to be given by the bin width
(which corresponds to $\sigma\approx 3.2/\sqrt{12}\approx 1$\,MeV), and is neglected. 
The narrow resonance requires four additional parameters: $M$, $\Gamma$, and the
product of helicities $A_{1/2}, A_{3/2}$ and the square root of the $N\eta'$ decay branching
fraction.

Different spin-parity combinations were tested: $J^P=1/2^\pm, 3/2^\pm, 5/2^\pm, 7/2^\pm$.
Fits with  $J^P=1/2^\pm$ and $J^P=7/2^\pm$ gave no improvement; the best fit was achieved
with quantum numbers $J^P=3/2^-$ but $5/2^-$ could also be possible. For the first two bins
in Fig.~\ref{asymmetry}, 
the beam asymmetry was calculated 
for 1\,MeV wide bins and then averaged. For the other bins, the beam asymmetries were calculated
for the central masses. The dashed curves in Figs.~\ref{asymmetry} show this fit.   
Assuming $J^P=3/2^-$, the fit returns
\begin{eqnarray}
M_{\eta'p}=1900\pm1\,{\rm MeV}; && \Gamma_{\eta'p} <3\,{\rm MeV}\,.
\end{eqnarray}

When the narrow resonance is included, the $\chi^2$ of the fit improves from 
120.3 to 59.9 for the 70 data points in the first five Mainz mass bins, or from 
29.5 to 11.7 for the 14 GRAAL data points. Due to the strongly rising phase space,
the narrow resonance entails a small $J^P=3/2^-$ amplitude extending over more than 10\,MeV. 
Note that the mass resolution is given by the photon beam energy and not by
the reconstruction of the final state. 

\begin{figure}[pt]
\vspace{4mm}
 \center
    \includegraphics[width=0.48\textwidth,height=0.35\textheight]{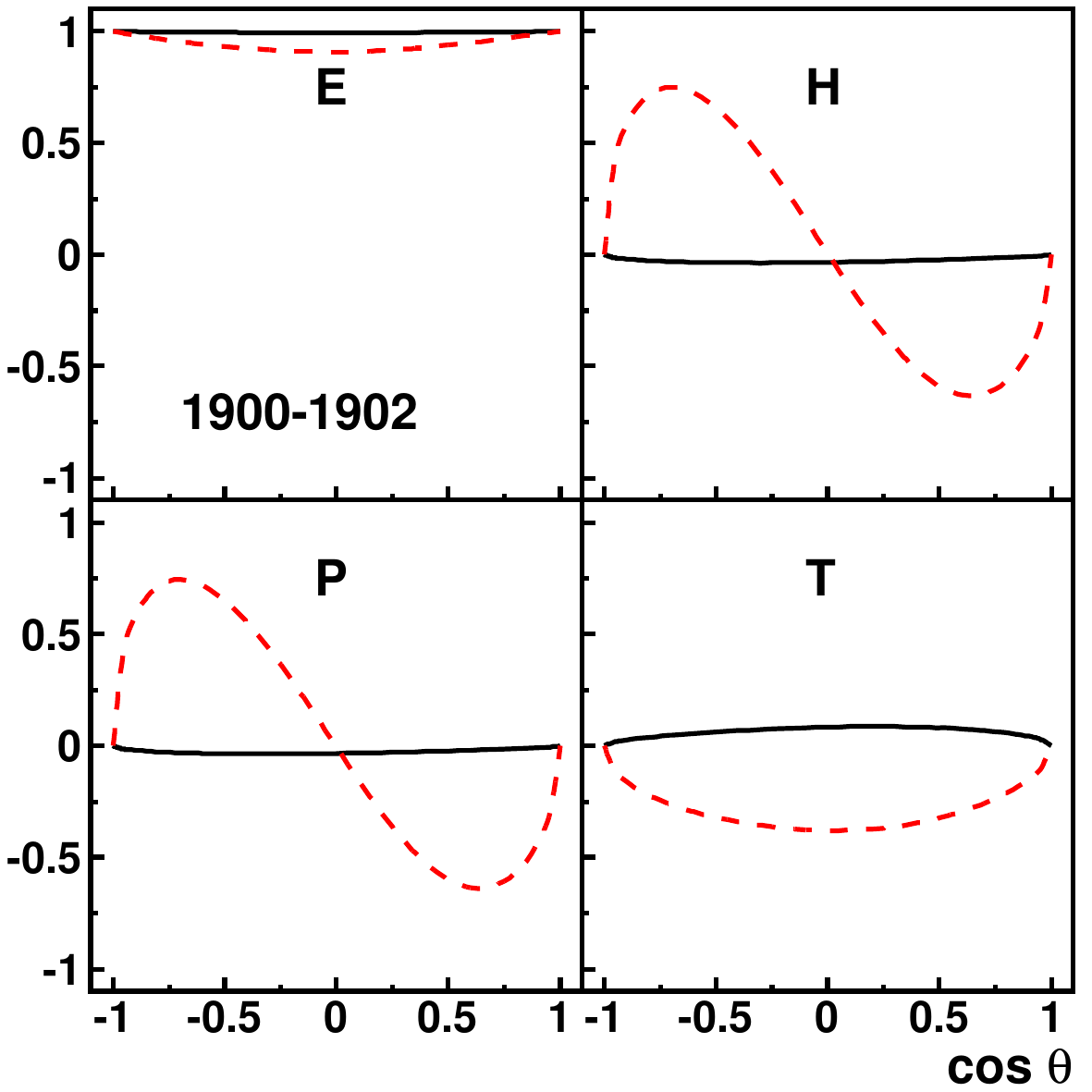}
  \caption{\label{EG}(Color online) Prediction for the double-polarization
	observables $E$, $G$, $P$ and $T$ for the 1901 - 1902\,MeV  mass window. 
	The solid (black) curve represents the main fit without a new narrow resonance, 
	the dashed (red) curve a fit represents the prediction for the case
when a narrow $\eta'p$ threshold resonance with spin-parity $J^P=3/2^-$ exists.
}
\end{figure}

The existence of such a narrow resonance in the $D$-wave is unexpected. 
(In the $S$-wave, a $N\eta'$ bound state just below the threshold is predicted
in the linear sigma model \cite{Sakai:2014zoa}). To trace the origin
of the narrow structure, we have excluded the MAMI-A2 data from the fit. If we impose 
the GRAAL resolution of $16$\,MeV (FWHM), the narrow width is confirmed. Decreasing 
the resolution leads to a larger natural width of the narrow resonance. Thus, the evidence for the
narrow width of the resonance rests mostly on the MAINZ $\gamma p\to \eta' p$ differential cross
section~\cite{Kashevarov:2017kqb}.

The GRAAL data suggest the possible existence of a $p\eta'$ threshold resonance; the MAMI-A2
differential cross sections support this conjecture. Hence we think that the search
for a narrow $N\eta'$ resonance should be continued with new data with high statistics and
precision. A measurement of further polarization observables might help to find an 
unambiguous answer. The observables for which predictions are made include those for 
several polarized-photon-beam/polarized-proton-target combinations.

Figure~\ref{EG} shows the predictions for the polarization observables $E$, $H$, $P$ and $T$
for a 2\,MeV mass region at the nominal mass of the possible $\eta' p$ resonance. The 
observable $E$ is the (normalized) difference between the meson photoproduction cross 
sections for helicity 1/2 and helicity 3/2. The observable $H$ is the correlation between 
linearly-polarized photons and transversely-polarized protons. The observable $P$ is the 
polarization of the outgoing proton with respect to the scattering plane. The observable $T$ 
is the asymmetry in the production cross section when the target proton polarization 
transverse to the incident photon beam is flipped. 

When a resonance is added to the fit, significant differences in one or more of these 
observables should arise. In particular, data for $P$ and $H$ should indicate the 
presence of a resonance (even if the resonance is narrow) if such a structure exists,
and predictions for those observables are relatively sensitive to the presence of that state.
However, we also recognize that experiments so near to threshold are very demanding,
especially if a putative resonance is so narrow that the polarization
observables can only show evidence for that resonance in a small mass window.

\begin{figure}[pt]
\center
      \includegraphics[width=0.48\textwidth,height=0.35\textheight]{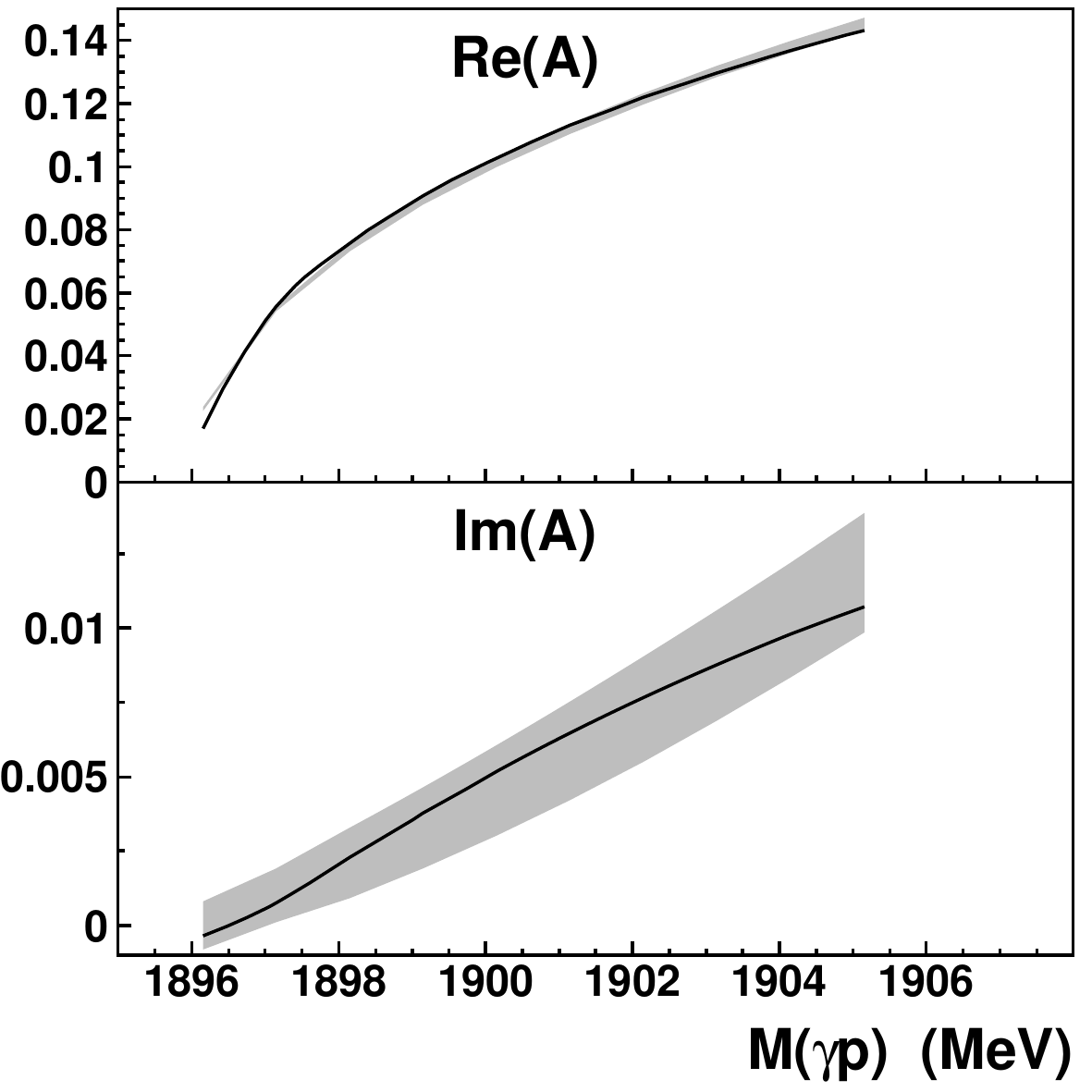}
  \caption{\label{swave}(Color online) Real part (upper panel) and imaginary part (lower panel) of the 
	$\eta' N\to \eta' N$ S-wave scattering amplitude $A$ (solid curve) using Eqn.~(\ref{eqscatt}). 
	The phase is rotated to have a vanishing imaginary
   part at the $\eta' p$ threshold. The error band is calculated from the variance of the 
	partial-wave analysis coefficients. 
}
\end{figure}

We now turn to the main objective of this paper: the determination of the $\eta' p$ scattering length.
The scattering length is given by the $J^P=1/2^-$- ($S_{11}$)-wave amplitude for $\eta' p$ production at 
the $\eta'$ threshold (corresponding to the wave in which the $\eta' p$ orbital angular 
momentum vanishes). As the primary solution, we use the {\it standard fit} without additional
narrow resonance. The upper and lower panels of Figure~\ref{swave} show the real and imaginary parts, respectively, of the $S_{11}$ partial wave in the low-energy region, from the threshold to 10\,MeV above. 
A fit to the squared amplitude with $\beta\cdot \sqrt{s-s_0}$ yields an offset of $1896.0\pm 1.0$\,MeV, 
fully consistent with 
the sum of proton and $\eta'$ mass: $1896.05\pm 0.06$\,MeV.

The results shown in Figs.~\ref{swave}a and b were fit with a function
\begin{eqnarray}
\label{eqscatt}
A&=&\frac{a\ k}{1\ -i\ k\ a\ +R\ k^2\ a\ /2\ + d\ k^4\ a} \\
k&=&\frac{\sqrt{(s-(M_p+M_{\eta'})^2)}}{4\ s}\nonumber
\end{eqnarray}
where $a=a_{p\eta'}$ is the $\eta' p$ scattering length, $R$ the interaction range
and $d$ a parameter representing higher-order terms. The fit to Figs.~\ref{swave}a and b give consistent
results for the modulus of the $\eta' p$ scattering length.

The $S_{11}$ amplitude has one arbitrary phase which cannot be defined from experiment. 
The phase could be defined as zero at the pion-production threshold or, alternatively, at 
the $\eta'$ production threshold. 

We fit the amplitude in the range from threshold to 10\,MeV above the threshold, 
the parameter $d$ was fixed to zero or was used as a free fit parameter. From these 
fits, we obtained a modulus of the scattering length in the range $0.367 > |a_{p\eta'}| > 0.344$\,fm with
a statistical error ($\sigma$) of less than $\pm 0.013$\,fm. 
Then we performed fits with an additional narrow resonance, again with $d=0$ and with $d$ 
as free fit parameter. In this case the fit returned values 
between $0.462 > |a_{p\eta'}| > 0.394$\,fm and a statistical error of $\pm 0.020$\,fm.  
The modulus of the $\eta' p$ scattering length depends only weakly
on the existence of a narrow $D$-wave resonance at the $\eta' p$ threshold.
Since we do not know if a narrow resonance
exists or not, we quote
\beq
|a_{p\eta'}| = (0.403\pm 0.020\pm 0.060)\,{\rm fm}
\eeq
as our final result, where the first uncertainty is statistical, the second one is nature. 
The phase relative to the $\eta'$ production threshold is
\beq
\delta =(1.5\pm 0.5)^\circ;   
\eeq
With this definition of the phase, the imaginary part of the amplitude nearly vanishes.
This scattering length would lead to a value for the real part of the $\eta'$ potential of $33\pm5$\,MeV
and a very small ($\approx 1$\,MeV) imaginary part, making the search for $\eta'$-nucleus
bound states particularly attractive.

The phase shift between the $\pi N$ and the $\eta' p$ threshold is rather stable. 
If the phase is defined relative to the pion-production threshold, we find
\beq
\delta =(87\pm 2)^\circ .
\eeq
These values are consistent with the numbers obtained at COSY~\cite{Czerwinski:2014yot}:
$\mathrm{Re}(a_{p\eta'}) =  0~\pm~0.43~\mathrm{fm}$ and
$\mathrm{Im}(a_{p\eta'}) = 0.37^{~+0.40}_{~-0.16}~\mathrm{fm}$.

The phase of about $90^\circ$ implies that the real part of the scattering 
length is small compared to imaginary part. The existence of a $\eta' p$
($S$-wave) bound state would require the real part to be larger than the 
imaginary part \cite{Haider:2015fea}. This value for the scattering length does not, 
however, exclude the existence of a $\eta' N$ bound state in a nuclear
environment or the existence of a $D$-wave $\eta' N$ resonance.

The range parameter is a complex number which remained essentially undetermined. We find 
$R=(3.2\pm2.1) +i(1.6\pm 1.2)$\,fm for fits with $d=0$ and $R=(10\pm 5) + i(5\pm 5)$\,fm 
when $d$ is varied freely. There was wide variation of the estimates for the value of 
(and uncertainty for) $d$, so no specific value for that quantity is quoted here.

\begin{figure}[pt]
 \center
      \includegraphics[width=0.4\textwidth,height=0.33\textwidth]{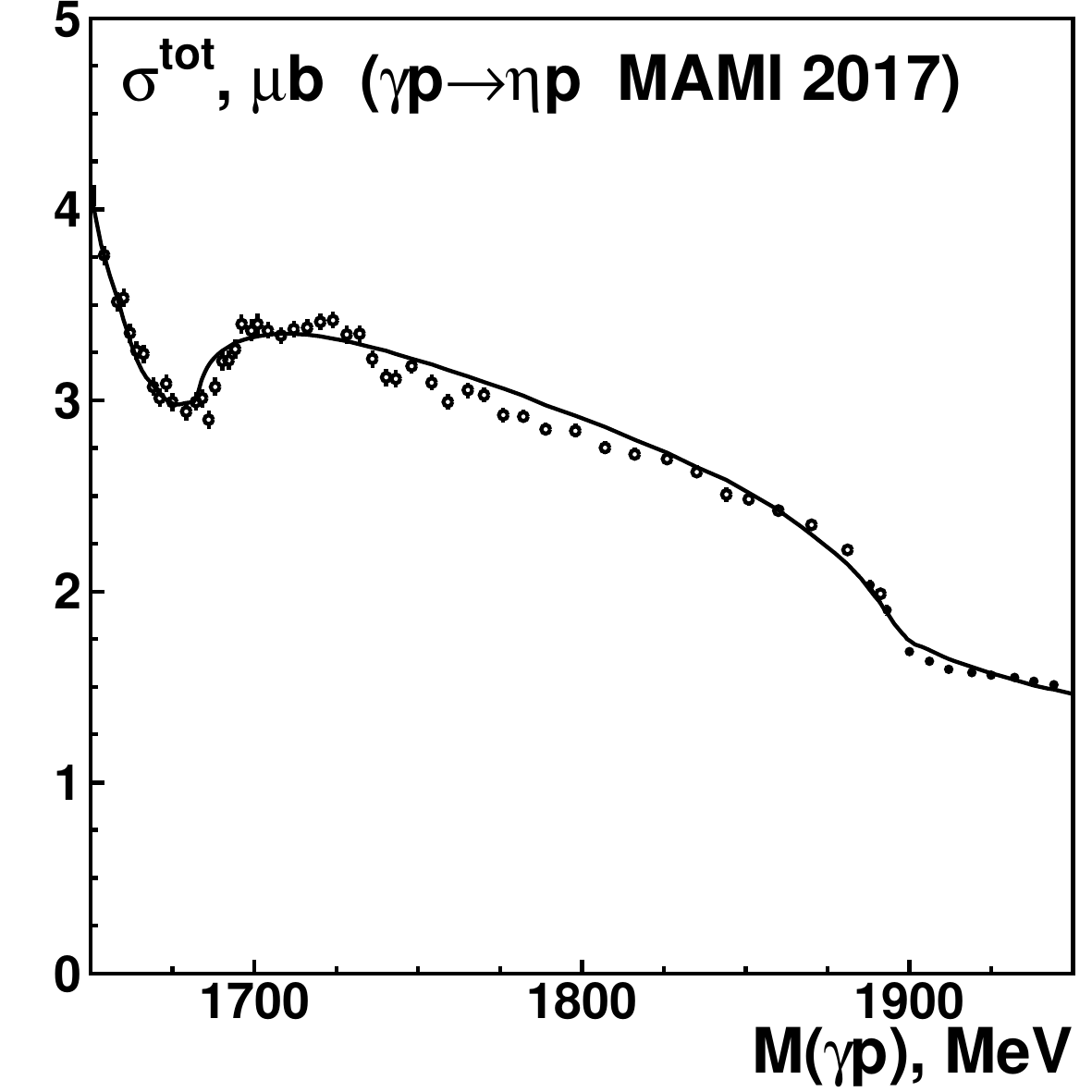}
   \caption{\label{etatot}(Color online) Total cross section for 
$\gamma p\to \eta p$ Ref.~\cite{Kashevarov:2017kqb}. The significant 
drop at the $\eta' p$ threshold near 1890 MeV indicates the presence of strong contributions from the
$N(1895)1/2^-$ resonance. The data shown come from two running periods. The cross sections
for the high-mass data above 1880\,MeV have been scaled with a factor 1.06.
}
\end{figure}

In all fits, $N\eta$ decay modes of most resonances are admitted. The opening of the
$\eta' p$ threshold can be seen in the data as pointed out in Ref.~\cite{Kashevarov:2017kqb}.
The data are shown in Fig.~\ref{etatot} and compared to the BnGa fit. The drop of the cross section 
at about 1890\,MeV is due to the opening of the $\eta'$ threshold and indicates
the presence of a resonance at about this mass with quantum number $J^P=1/2^-$. 

Summarizing, we have studied the photoproduction reaction $\gamma p\to \eta' p$. The GRAAL
data on the beam asymmetry for this reaction suggest the possible existence of a narrow $\eta' p$
resonance at $1900\pm 1$\,MeV and a width of less than $3$\,MeV. The $\eta' p$ scattering
length has been determined. Its magnitude is found to be 
$|a_{p\eta'}|=(0.403\pm 0.020\pm0.0600)$\,fm, its phase
relative to the $\pi N$ threshold is $\delta=(87\pm2)^\circ$. This value does not depend 
on the existence or not of the narrow $\eta' p$ resonance. The range of the interaction
could not be determined with a reasonable accuracy.
\vspace{2mm}

{\it\small
We would like to thank V. Metag for illuminating discussions on $\eta'$ interactions in nuclei.
This work was supported by the Deutsche Forschungsgemeinschaft (SFB/TR110) and by the
RSF grant 16-12-10267. Work at Arizona State University was supported by the U.~S.~National
Science Foundation under award PHY-1306737. This material is in part based upon work
supported by the U.S. Department of Energy, Office of Science, Office of Nuclear Physics
under contract DE-AC05-06OR23177.
}


\vspace{2mm}


\begin{thebibliography}{99}
\bibitem{Coffman:1988ve}
  D.~Coffman {\it et al.} [MARK-III Collaboration],
  Phys.\ Rev.\ D {\bf 38}, 2695 (1988).

\bibitem{Feldmann:1998vh}
  T.~Feldmann, P.~Kroll and B.~Stech,
  Phys.\ Rev.\ D {\bf 58}, 114006 (1998).

\bibitem{Escribano:2008rq}
  R.~Escribano,
  Eur.\ Phys.\ J.\ C {\bf 65}, 467 (2010).

	\bibitem{Cao:2012nj}
  F.~G.~Cao,
  Phys.\ Rev.\ D {\bf 85}, 057501 (2012).

\bibitem{Aaij:2014jna}
  R.~Aaij {\it et al.} [LHCb Collaboration],
  JHEP {\bf 1501}, 024 (2015).

\bibitem{Harland-Lang:2017mse}
  L.~A.~Harland-Lang, V.~A.~Khoze, M.~G.~Ryskin and A.~G.~Shuvaev,
  Phys.\ Lett.\ B {\bf 770}, 88 (2017).

\bibitem{tHooft:1976snw}
  G.~'t Hooft,
  Phys.\ Rev.\ D {\bf 14}, 3432 (1976)
  Erratum: [Phys.\ Rev.\ D {\bf 18}, 2199 (1978)].

\bibitem{Bernard:1988db}
  V.~Bernard and U.~G.~Meissner,
  Nucl.\ Phys.\ A {\bf 489}, 647 (1988).

\bibitem{Brown:1991kk}
  G.~E.~Brown and M.~Rho,
  Phys.\ Rev.\ Lett.\  {\bf 66}, 2720 (1991).

\bibitem{Hatsuda:1991ez}
  T.~Hatsuda and S.~H.~Lee,
  Phys.\ Rev.\ C {\bf 46}, no. 1, R34 (1992).

\bibitem{Sekihara:2017xkz} 
  T.~Sekihara, H.~Fujioka and T.~Ishikawa,
  ``Possible $\eta' d$ bound state and its $s$-channel formation in the $\gamma d \to \eta d$ reaction,''
  arXiv:1712.06257 [nucl-th].
	
\bibitem{Nanova:2013fxl}
  M.~Nanova {\it et al.} [CBELSA/TAPS Collaboration],
  Phys.\ Lett.\ B {\bf 727}, 417 (2013).

\bibitem{Nanova:2016cyn}
  M.~Nanova {\it et al.} [CBELSA/TAPS Collaboration],
  Phys.\ Rev.\ C {\bf 94}, no. 2, 025205 (2016).

 \bibitem{Metag:2017yuh} 
  V.~Metag, M.~Nanova and E.~Y.~Paryev,
  Prog.\ Part.\ Nucl.\ Phys.\  {\bf 97}, 199 (2017).
	
	\bibitem{Nanova:2018tbd}
  M.~Nanova {\it et al.} [CBELSA/TAPS Collaboration],
  ``The ${\eta}$'-Carbon potential at low meson momenta,''
  in preparation.

	\bibitem{Czerwinski:2014yot}
  E.~Czerwinski {\it et al.},
  Phys.\ Rev.\ Lett.\  {\bf 113}, 062004 (2014).

\bibitem{Sandri:2014nqz}
  P.~Levi Sandri {\it et al.},
  Eur.\ Phys.\ J.\ A {\bf 51}, no. 7, 77 (2015).

\bibitem{Collins:2017sgu}
  P.~Collins {\it et al.} [CLAS Collaboration],
  Phys.\ Lett.\ B {\bf 771}, 213 (2017).

\bibitem{Kashevarov:2017kqb}
  V.~L.~Kashevarov {\it et al.},
  Phys.\ Rev.\ Lett.\  {\bf 118}, no. 21, 212001 (2017).

 \bibitem{Anisovich:2017tbd}
A.~V.~Anisovich {\it et al.},
``$N^*\to N \eta^\prime$ decays from photoproduction of $\eta^\prime$-mesons off protons'',
submitted to Physics Letters B.

	\bibitem{Crede:2009zzb}
  V.~Crede {\it et al.} [CBELSA/TAPS Collaboration],
  Phys.\ Rev.\ C {\bf 80}, 055202 (2009).

\bibitem{Williams:2009yj}
  M.~Williams {\it et al.} [CLAS Collaboration],
  Phys.\ Rev.\ C {\bf 80}, 045213 (2009).

 \bibitem{Arndt:2006bf}
  R.~A.~Arndt, W.~J.~Briscoe, I.~I.~Strakovsky and R.~L.~Workman,
  Phys.\ Rev.\ C {\bf 74}, 045205 (2006).

\bibitem{Zhong:2011ht} 
  X.~H.~Zhong and Q.~Zhao,
  Phys.\ Rev.\ C {\bf 84}, 065204 (2011).
	
\bibitem{Huang:2012xj} 
  F.~Huang, H.~Haberzettl and K.~Nakayama,
  Phys.\ Rev.\ C {\bf 87}, 054004 (2013)	

\bibitem{Sakai:2016boo} 
  S.~Sakai, A.~Hosaka and H.~Nagahiro,
  Phys.\ Rev.\ C {\bf 95}, no. 4, 045206 (2017).
	
\bibitem{Tryasuchev:2017amh} 
  V.~A.~Tryasuchev, A.~G.~Kondratyeva and A.~A.~Kiziridi,
  Russ.\ Phys.\ J.\  {\bf 60}, no. 5, 782 (2017).	
	
\bibitem{Jaegle:2010jg} 
  I.~Jaegle {\it et al.} [CBELSA/TAPS Collaboration],
  Eur.\ Phys.\ J.\ A {\bf 47}, 11 (2011).
	
	\bibitem{Klaja:2010vy} 
  P.~Klaja {\it et al.},
  Phys.\ Lett.\ B {\bf 684}, 11 (2010).

\bibitem{Sakai:2014zoa} 
  S.~Sakai and D.~Jido,
  Hyperfine Interact.\  {\bf 234}, no. 1-3, 71 (2015).
	
	\bibitem{Green:2004tj}
  A.~M.~Green and S.~Wycech,
  Phys.\ Rev.\ C {\bf 71}, 014001 (2005)
  Erratum: [Phys.\ Rev.\ C {\bf 72}, 029902 (2005)].

\bibitem{Haider:2015fea} 
  Q.~Haider and L.~C.~Liu,
  Int.\ J.\ Mod.\ Phys.\ E {\bf 24}, no. 10, 1530009 (2015).
\end{thebibliography}
\end{document}